\pdfoutput=1
\documentclass{article}

\usepackage{CJKutf8}
\usepackage{PRIMEarxiv}
\usepackage[utf8]{inputenc} 
\usepackage[T1]{fontenc}    
\usepackage{hyperref}       
\usepackage{url}            
\usepackage{booktabs}       
\usepackage{amsfonts}       
\usepackage{nicefrac}       
\usepackage{microtype}      
\usepackage{lipsum}
\usepackage{fancyhdr}       
\usepackage{graphicx}       
\usepackage{tabularx}
\usepackage{float}  
\usepackage{subfigure}  
\usepackage{cite}
\usepackage{svg}
\graphicspath{{media/}}     
\usepackage{svg}
\svgsetup{
inkscapepath=i/svg-inkscape/
}
\svgpath{{media/}}
\usepackage[normalem]{ulem}
\useunder{\uline}{\ul}{}
\title{Physics Sensor Based Deep Learning Fall Detection System
\thanks{\textit{\underline{Citation}}: 
\textbf{Corresponding author: lyj@nwpu.edu.cn}} 
}
\begin{document}

\author{
  Zeyuan Qu, Tiange Huang, Yuxin Ji, Yongjun Li* \\
  School of Computer, Northwestern Polytechnical University, Xi'an, Shaanxi 710072, China. \\
}
\maketitle

\begin{abstract}
Fall detection based on embedded sensor is a practical and popular research direction in recent years. In terms of a specific application: fall detection methods based upon physics sensors such as [gyroscope and accelerator] have been exploited using traditional hand crafted features and feed them in machine learning models like Markov chain or just threshold based classification methods. In this paper, we build a complete system named TSFallDetect including data receiving device based on embedded sensor, mobile deep-learning model deploying platform, and a simple server, which will be used to gather models and data for future expansion. On the other hand, we exploit the sequential deep-learning methods to address this falling motion prediction problem based on data collected by inertial and film pressure sensors. We make a empirical study based on existing datasets and our datasets collected from our system separately, which shows that the deep-learning model has more potential advantage than other traditional methods, and we proposed a new deep-learning model based on the time series data to predict the fall, and it may be superior to other sequential models in this particular field.
\end{abstract}


\section{Introduction}


According to the World Health Organization, the number and proportion of people aged 60 years and older in the population is increasing. In 2019, the number of people aged 60 years and older was 1 billion. This number will increase to 1.4 billion by 2030 and 2.1 billion by 2050. With the decline of physical function, the probability of falling in the elderly increases \cite{who2023}. The World Health Organization estimates that 28\% to 35\% of older adults (>=65 years) have at least one fall per year. If you can't get up after falling on the ground for more than 1 hour, this is called \emph{long lye} \cite{2007Falls}. \emph{Long lye} is dangerous for the elderly, which can lead to illness and even death. In China, accidental falls are already the leading cause of fatal and non-fatal injuries among people aged 65. How to prevent falls in time and conduct effective early warning or notification after falls to avoid subsequent injuries is a technical problem that needs further research.

For many years, research activity has focused on two main areas related to falls, namely prevention and detection \cite{2014Survey}. These two areas are related to each other \cite{Kabalan2017From} because effective fall prevention requires accurate fall detection. In terms of fall detection methods, several methods are proposed, which can be mainly divided into wearable and non-wearable solutions.




Non-wearable solutions primarily utilize environmental sensors for fall detection. However, regardless of any specific sensing technology used, there are some limitations to environmental sensor-based systems. They are generally suitable for indoor spaces, and in most cases, they are more expensive and require more computing resources. In addition, the vision sensor may be affected by light conditions and field of view.



The advantages of wearable devices are more prominent and suitable in terms of calculation and development cost. In addition, when a wearable sensor is used to detect a fall, the position of the sensor on the human body will affect the [detection ability] of the system \cite{2016Investigation}. Barshan et al. \cite{2014Detecting} fitted the six wireless sensor units tightly with special straps to the subjects' head, chest, waist, right wrist, right thigh, and right ankle. And the use of waist sensor units \cite{2016An} generally results in the best accuracy of fall detection. The foot sensor is a special type of wearable sensor that can perform accurate motion capture and provide a better human wear experience. Once the design is complete, there is no need for any external transceivers in indoor and outdoor environments. This way, users can walk comfortably without being affected by the fall detection system. Based on the above analysis, we designed a fall detection system based on sensors deployed in the foot.

We built a set of fall detection system, including an embedded sensing hardware device with data collection and transmission functions, a mobile client with data receiving, running fall detection model and data preservation functions, and an experimental server with cloud storage functions. On the basis of this, we collected fall data (pressure acceleration, foot angular speed, etc., when falling or walking normally). Meanwhile, we designed a fall detection model FallSeqTCN. Based on the sequence data, we trained the model and got a fall classifier, and deployed it in the Mobile Client. At the same time, the data generated in the process of using the APP will be uploaded to the server built by us as a historical data backup.

In the specific model construction, we propose a simple binary classification model for falls and non-falls based on TCN, FallSeqTCN, which is a time series prediction network based on extended convolution. It has a special structure of expansive causal convolution, which can effectively capture a long period of historical information before the fall. At the same time, we add residual structure to the network to deal with the too deep network structure, which also makes our model have a good generalization ability. The results show that our fall classifier is better. Finally, we deploy the whole set of models on the system and realize the whole set of fall detection system with practical significance.


In short, our deep learning fall detection system based on physical sensor has demonstrated reliable results in detecting falls accurately with no false alarms. Our system can potentially be used as an effective and low-cost fall detection system, enabling healthcare professionals to respond early and quickly when incidents occur. In summary, our contributions are as follows.

\begin{itemize}
  \item[1)]
  We designed and constructed a set of human motion data intelligent sensing system. The sensor data acquisition device adopts a wearable scheme, which can transmit raw data in real time. The mobile phone application can not only analyze the motion parameters in real time, but also display, store and visualize the parsed data. Finally, we stored the uploaded data and updated the model through the Minio server. The overall system can not only analyze the state of human movement without affecting people's normal actions, but also detect falls at a higher resolution and early response during an accident.
  \item[2)]
  We propose a simple binary classification model for falls and non-falls based on TCN, FallSeqTCN. It has a special extended causal convolution and residual structure, which not only allows us to make more efficient use of the captured long historical information before the fall, but also enables our model to have good generalization ability.
  \item[3)]
  We verify the reliability of FallSeqTCN model on two public data sets and conduct training tests on the actual data sets. By analyzing the data and adjusting the data set, we get a reliable fall classifier, which is very suitable for the actual fall detection equipment, and there is no missing judgment.
\end{itemize}

The structure of this article is as follows. In the second section, we summarize the existing implementation methods and literature of the fall detection system, analyze the existing problems and put forward the corresponding solutions. At the same time we make a brief description of our system in this section. In the third section, we mainly describe the specific implementation methods of each module, including hardware platform, APP and server. The fourth section introduces deep learning model used in this system. The fifth section discusses the data collection and evaluation results of the model. Finally, the sixth section summarizes the main work.






\section{Related works}

\subsection{Existing related detection systems}

Many studies have proposed the use of vision sensors as a solution. Frequently seen vision sensors include depth camera sensors \cite{2014A}\cite{2017Fall} and radar sensors \cite{2017Radar}. Depth camera sensors can provide three-dimensional information about the scene, including distance and depth data, to better understand the state of objects and human bodies in the environment. 
In addition to this, several other solutions have been proposed, such as acoustic signal-based solutions \cite{2015An} and smart flooring embedded with pressure-sensitive fibers \cite{2016Floor}.

On the contrary, with the support of the development of Microelectromechanical systems (MEMS), the advantages of wearable devices are more prominent and suitable in terms of [calculation and development cost]. For these reasons, they provide a viable means for scalable health monitoring of the elderly \cite{2017Wearables}\cite{Mukhopadhyay2014Wearable}.

Although the use of waist sensor units \cite{2016An} generally results in the best accuracy of fall detection, El-Bendary et al., however, emphasize that the use of wearable sensors may affect [user acceptance] \cite{2013Fall}, especially when the sensor or its location is sensitive.
Previous studies have explored the potential of using machine-assisted fall detection in a variety of domains, including vision-based fall detection using camera sensors \cite{2014A}\cite{2017Fall}\cite{9471869}\cite{vicnn63}\cite{vicnn64}\cite{3DSkeleton80}, fall detection based accelerometer using smartphones \cite{2020A}, and inertial measurement unit data based fall detection methods\cite{CasilariPrez2020ASO77}\cite{cnnlstmIoTEnabled99}\cite{Torti2018EmbeddedRF105}\cite{Theodoridis2018HumanFD106}. 

For vision based fall detection, the utilization of Convolutional Neural Networks (CNNs) was prominent. Two notable studies\cite{vicnn63}\cite{vicnn64}, incorporated depth camera data as a key input. Paper\cite{vicnn63} introduced a fall detection system employing video frames captured by Kinect RGB depth cameras. CNNs were used to distinguish between Activities of Daily Living (ADL) and fall events. The dataset consisted of 21,499 images collected from various individuals in different indoor environments. The dataset was divided into training and testing sets in a 73-27 ratio, yielding an overall accuracy of 74\%. In the other paper\cite{vicnn63} focused on extracting human body shape deformity features using CNNs directly on frame images. The URFD dataset was used, and the system's performance was evaluated through 10-fold cross-validation, achieving an average sensitivity of 100\%, specificity, and accuracy of 99.98\%. However, it's worth noting that the system's performance may be affected by limited background and foreground variations in the dataset. Additionally, a novel approach, MyNet1D-D, proposed by Tsai and Hsu \cite{3DSkeleton80}, introduced a robust one-dimensional CNN architecture for transforming depth image data into skeleton information, emphasizing computational efficiency and suitability for embedded systems.

In the realm of sensor-based fall detection, Casilari et al. Paper\cite{CasilariPrez2020ASO77} presented a fall detection system based on deep CNNs. Their system identified fall events by recognizing patterns in three-axis accelerometer data. It incorporated an extensive dataset that encompassed up to 14 publicly available datasets, including MobiAct, SisFall, MobiFall, UniMiB SHAR, and UP-Fall. Meanwhile, paper\cite{cnnlstmIoTEnabled99} harnessed the power of CNNs and Long Short-Term Memory networks (LSTMs) to perform feature extraction on time-series accelerometer data. This approach outperformed a combination of Support Vector Machine (SVM) and CNN-based methods. The method introduced in the paper\cite{Torti2018EmbeddedRF105} introduced a method for implementing fall detection with Recurrent Neural Network (RNN) architectures, including LSTMs, making it compatible with microcontroller units (MCUs) equipped with three-axis accelerometers. This approach builds upon previous work by Theodoridis et al. \cite{Theodoridis2018HumanFD106} on RNN-based fall detection, demonstrating the capability of handling and encoding sequential data obtained from body-worn accelerometers.

\subsection{Problems}
Based on the analysis of the existing research, we present the following questions.
\begin{itemize}
    \item[1)]
    Fall detection solutions have suffered from a lack of reliable fall detection algorithms that can accurately and reliably detect falls, particularly those involving dynamic movement changes and combinations of movements. This is due in part to the difficulty of modelling static poses with static models, which are insufficient to capture changes in dynamic body configurations and movements during falls.
    \item[2)]
    Traditional wearable sensors, such as waist sensors and leg sensors, tend to cause great discomfort when users wear them, which has affected the normal life of users. In addition, the strange eyes brought by different clothes are also a reason why users are not willing to use.
    \item[3)]
    Deep learning models are sensitive to noise and undesired signals, which may lead to biases and inaccuracies in the detection of falls. Traditional sensors and threshold-based systems may be prone to a variety of noise sources, such as extraneous vibrations and other distorting effects, leading to inaccurate detection.
\end{itemize}

\section{System construction}
\subsection{Overview of our system}

\begin{figure}[htb]
    \centering
   \includegraphics[width=1\linewidth]{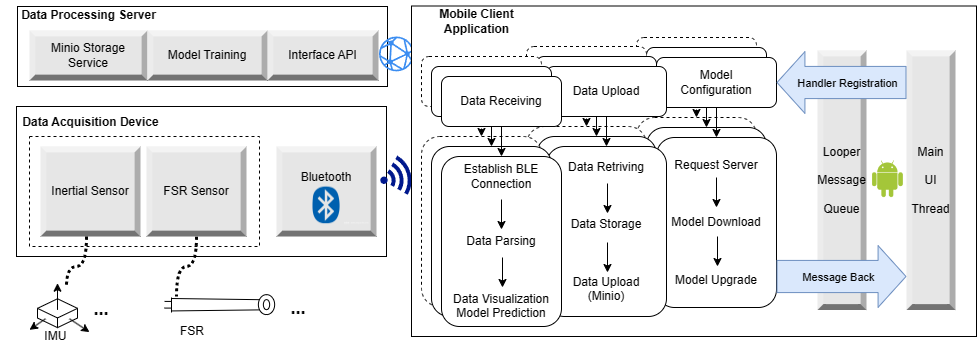}
   \caption{System Structure}
   \label{fig:overview}
\end{figure}

In response to challenges discussed above, we propose a physical sensor-based deep learning fall detection system and make a prototype of the system. You can see this in the \ref{fig:overview}. It mainly consists of three parts: data acquisition device, mobile client APP and an experimental server. The sensor data acquisition device mainly initializes the sensor and transmits the original data in real time. The mobile application is to realize the mobile phone BLE receiving library, UI interactive viewing and operation of connected devices. At the same time, we can analyze the motion parameters transmitted by the hardware data acquisition system to the smart phone in real time on the APP, and store and visualize the parsed data. The Minio server is used to save data and update models.

\begin{itemize}
    \item [1)]
    \textbf{Data Acquisition Device} We use physical sensors including accelerators, gyroscopes and pressure sensors to measure the pressure, acceleration and orientation generated during motion. The data sensing device uses the data collected by these sensors to send the data to the mobile phone APP through Bluetooth for decoding. 
    \item [2)]
    \textbf{Mobile Client Application} The mobile APP is based on the Android system platform. The main thread is responsible for registering the threads executing specific functions into the corresponding thread pool (queue), which mainly includes data receiving, visualization and model prediction pipelines, data uploading and model downloading sub-pipelines.
    \item [3)]
    \textbf{Data processing server} We see it as a data shelter, mainly used to store historical movement data generated by users over time.
\end{itemize}

\label{sec:ideas}

\subsection{Data Acquisition Device}

\begin{figure}[htb]
  \centering
  \includegraphics[width=0.5\linewidth]{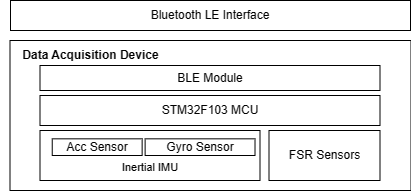}
  \caption{Data Acquisition Device}
  \label{fig:structure2}
\end{figure}

In view of the data uniformity and inaccuracy brought by a single sensor, we integrate multiple sensors based on STM32F103 MCU platform, including pressure sensor (FSR), voltage conversion module, acceleration, angular velocity and gyro sensor (IMU901) and Bluetooth communication module (ATK-BLE) . The overall prototype system architecture design is shown in figure \ref{fig:structure2}. The data acquisition system is designed as two independent data sources, each deployed on the left and right foot, which is taken into account the user experience. 

\begin{figure}[htb]
  \centering
  \includegraphics[width=0.5\linewidth]{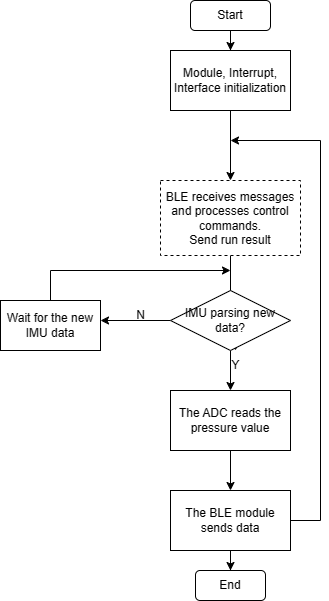}
  \caption{Data Acquisition Interaction}
  \label{fig:flow}
\end{figure}

\begin{figure}[htb]
  \centering
  \includegraphics[width=0.5\linewidth]{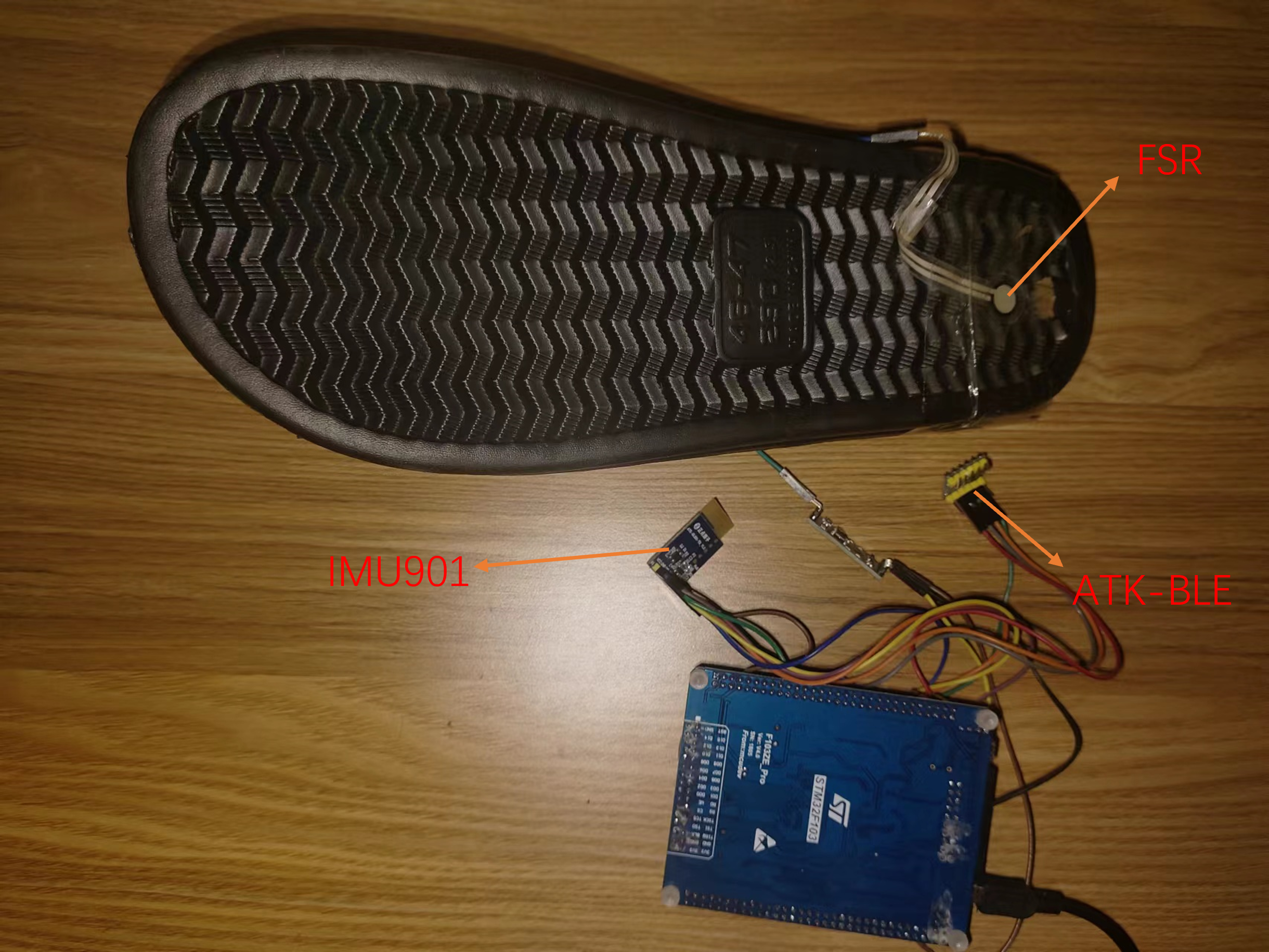}
  \caption{Our dataset obtained using this foot sensor group}
  \label{fig:device}
\end{figure}

See the figure \ref{fig:device}. The system uses Bluetooth protocol as an interface to communicate instructions and exchange data between embedded sensors platform and Android mobile platform.
The STM32 chip platform's firmware can communicate with PC through COM port for serial debugging and writing. The following is a brief introduction to the development platform of the data acquisition system, as well as each component. STM32 MCU adopts ARM architecture. In figure \ref{fig:flow}, the main loop of the program includes initialization and setting of each module, including GPIO initialization, timer interrupt, UART communication frequency convention, sending and receiving specific instructions for communication, and analyzing data structure serialization. Read the values of pressure, acceleration, angular velocity and altitude angles, and send the timing data to the mobile phone through the Bluetooth module for further data collection and processing.

Considering the compatibility of the model with the device, we chose a similar sensor on the later wearable sensor. The acquired signals include plantar pressure values, acceleration values, angular velocity values, and azimuth values. The embedded device is worn on the subject's left and right feet, and the sensor signals of the subject in the normal walking and falling state including forward fall and left fall are collected. The sampling frequency is 18HZ, and the signal collected each time includes 20 features of plantar pressure values, acceleration values, angular velocity values and azimuth on the left and right feet, constituting the initial time series dataset. In the same experimental environment, the subjects carried out 11 normal walks, 10 forward falls, and 5 left falls for a total of 26 experiments, which were sorted into 26 TXT files. The 20 characteristic values collected in each experiment are shown in Table \ref{tab:data}.

\begin{table}[htbp]
\centering
\caption{Signal characteristics acquired by embedded devices}
\begin{tabularx}{\textwidth}{c X c X}
\hline
Feature& Meaning& Feature& Meaning\\
\hline
l\_voltage\_ao & The voltage value of the left foot& r\_voltage\_ao & The voltage value of the right foot \\
L\_attitude\_roll & Azimuth around the x-axis on the left foot& r\_attitude\_roll & Azimuth around the x-axis on the right foot \\
L\_attitude\_pitch & Azimuth around the y-axis on the left foot& r\_attitude\_roll & Azimuth around the y-axis on the right foot \\
L\_attitude\_yaw & azimuth around the z-axis on the left foot& r\_attitude\_roll & azimuth around the z-axis on the right foot \\
L\_acc\_x & acceleration value on the x-axis on the left foot& r\_acc\_x & acceleration value on the x-axis on the right foot \\
L\_acc\_y & acceleration value on the y-axis on the left foot& r\_acc\_y & acceleration value on the y-axis on the right foot \\
L\_acc\_z & acceleration value on the z-axis on the left foot& r\_acc\_z & acceleration value on the z-axis on the right foot \\
L\_gyro\_x & The angular velocity value on the x-axis on the left foot& r\_gyro\_x & The angular velocity value on the x-axis on the right foot \\
L\_gyro\_y & The angular velocity value on the y-axis on the left foot& r\_gyro\_y & The angular velocity value on the y-axis on the right foot \\
L\_gyro\_z & The angular velocity value on the z-axis on the left foot& r\_gyro\_z & The angular velocity value on the z-axis on the right foot \\
\hline
\label{tab:data}
\end{tabularx}

\end{table}

\begin{figure}[htbp]
    \centering
    \includegraphics[width=0.5\linewidth]{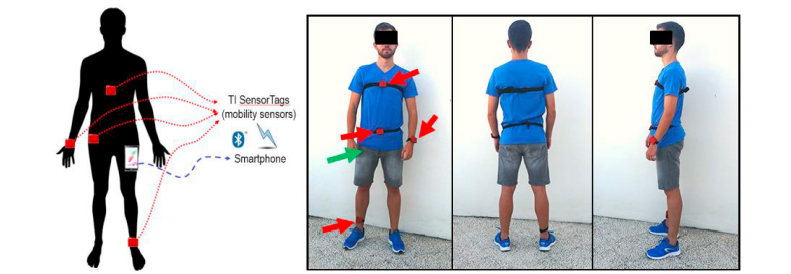}
    \caption{Basic architecture of the system}
    \label{fig:uma}
\end{figure}

Here we also introduce two similar public data sets. UMAFall is a new dataset of movement traces acquired through the systematic emulation of a set of predefined ADLs (Activities of Daily Life) and falls, as shown in figure \ref{fig:uma}. In opposition to other existing databases for FDSs, which only include the signals captured by one or two sensing points, the testbed deployed for the generation of UMAFall dataset incorporated five wearable sensing points, which were located on five different points of the body of the participants that developed the movements. We trained and tested the model on this.

\begin{figure}[htbp]
    \centering
    \includegraphics[width=0.5\linewidth]{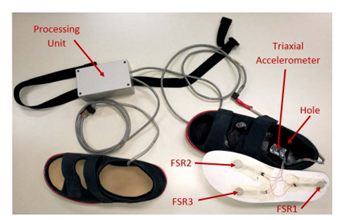}
    \caption{The TST FB4FD dataset obtained using the foot sensor group}
    \label{fig:tst}
\end{figure}

The TST FB4FD dataset was measured by a smart shoe, as shown in figure \ref{fig:tst}, which is equipped with three Force Sensing Resistors (FSR) and a three-axis accelerometer, as well as a processing unit board. The latter analyzes the gait cycle phase, distinguishes between falls and non-falls, and transmits data remotely.



\subsection{Mobile Client Application}

\begin{figure}[htbp]
    \centering
    \includegraphics[width=0.5\linewidth]{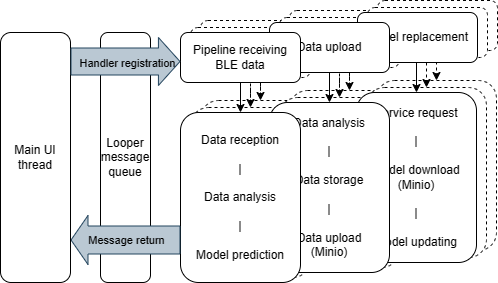}
    \caption{The framework diagram of APP design}
    \label{fig:appw}
\end{figure}

The concurrency problem caused by multi-device connection must be considered in the process of software design, and the Message-handler mechanism provided by the Android system platform can handle this problem well. Message represents an action or a sequence of actions, and each message has a specific target Handler when it is added to the message queue. Handler is the actual handler of the message. Handler allows Message and anonymous function objects associated with a thread's message queue to be sent and processed. Each Handler instance is associated with a message queue for a thread. Creating a Handler binds the Handler to a Looper. It can pass messages and anonymous function objects to Looper's message queue and execute them in Looper's thread. Handler has two main uses: (1) Schedule messages and anonymous function objects for execution at some point in the future; (2) Execute the action on a different thread.

The figure \ref{fig:appw} shows the process architecture of this design. The mobile APP we designed is based on the Android system platform, and the main thread is responsible for registering the thread executing specific functions into the corresponding thread pool (queue). There are mainly data receiving, visualization and model prediction pipelines, data upload and model download sub-pipelines. When the corresponding tasks are completed, the thread will encapsulate the results into Message and return to the main thread by Handler. Model prediction is supported by the Tensorflow Lite library.

\begin{figure}[!h]
    \centering
    \includegraphics[width=0.5\linewidth]{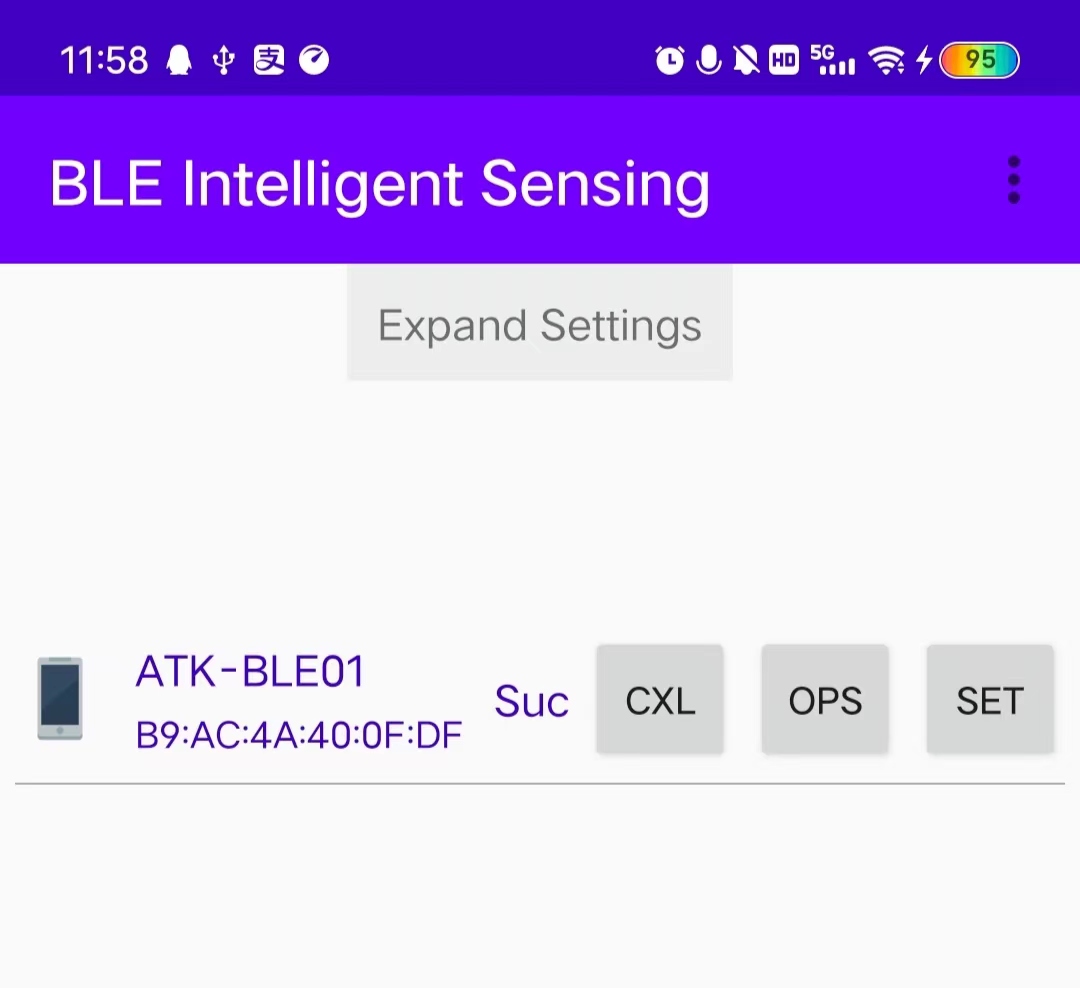}
    \caption{Device connection}
    \label{fig:app2}
\end{figure}

In order to facilitate the user's visual operation, we try our best to ensure the simplicity of the software, which is achieved on the basis of ensuring the integrity of fall detection and early warning functions.
The first is the ease of device connection, as shown in Figure \ref{fig:app2}, and users simply click the Connect button. 
[This is because after implementing the BLE connection library, we performed a registration scan connection callback and put the scanned devices into the list view adapter.]

\begin{figure}[!h]
    \centering
    \includegraphics[width=0.5\linewidth]{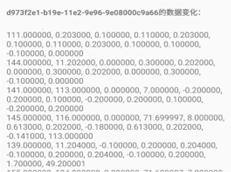}
    \caption{Data storage}
    \label{fig:app3}
\end{figure}

The form of data is also a problem that we focus on. On the one hand, our data should be convenient for subsequent model training, and on the other hand, these data can clearly describe the current motion state of users. Although real-time data reception is realized through the Bluetooth connection library, the Bluetooth connection library always receives byte fragments originally due to the byte limitation of BLE sending data. Proper data parsing and serialization became the top priority, and we implemented data processing for the data sink class.
The data sink class maintains an internal, concurrently secure string cache queue, lines\_queue, for parsed rows of data. The receiveData method accepts the raw byte type as a parameter, and is used to continuously concatenate the received byte fragments into the string cache, and truncate the byte fragments at the newline when the byte fragments contain newlines. The first half of the byte fragments is concatenated into the global string cache queue after the string cache is concatenated, and the second half is used to initialize the string cache. The parseData method uses an asynchronous callback scheme, when lines\_queue is not empty, the first line string (a set of raw data) is removed from the queue, and parsed into an internally defined BleUartData type, and marked isParsed as true, passed to a callback function initialized in the real application. It is used to perform the next operation on the parsed data. The data sink class provides an asynchronous callback function Interface after the data is parsed. We initialized the data parser in the onCreate function of the application interface portion of the code, and placed the receiveData and parseData calls in the main thread queue after successfully setting Notify to get the new raw data bytes. Finally, we set the file saving path and immediately pass in a sequence of bytes to save the data to the file. After testing, the frequency of data collection can be 12-80Hz. In the figure \ref{fig:app3}, we can see the storage format of the data.

To keep the interface simple, we use the open source ECharts library for data visualization.When the returned call data obtains the parsed data object, it displays the data in String format on the interface, and uses the parsed data tensor as input to run the deep learning model to obtain the motion state results corresponding to the motion data. These results can be visualized via EChartsView or used for motion state alarms.

In addition, to ensure that the model runs efficiently on devices with limited compute and memory resources, and to enable devices to run machine learning models offline, we use Tensorflow Lite as a set of tools to implement end-to-end workflows. TensorFlow Lite can be seen as consisting of two main parts: a converter that compresses and optimizes the model, converting it to.tflite format; a set of interpreters for various runtimes.
[Tensorflow Lite] 





\subsection{Data Processing Server}

\begin{figure}[htb]
    \centering
    \includegraphics[width=0.5\linewidth]{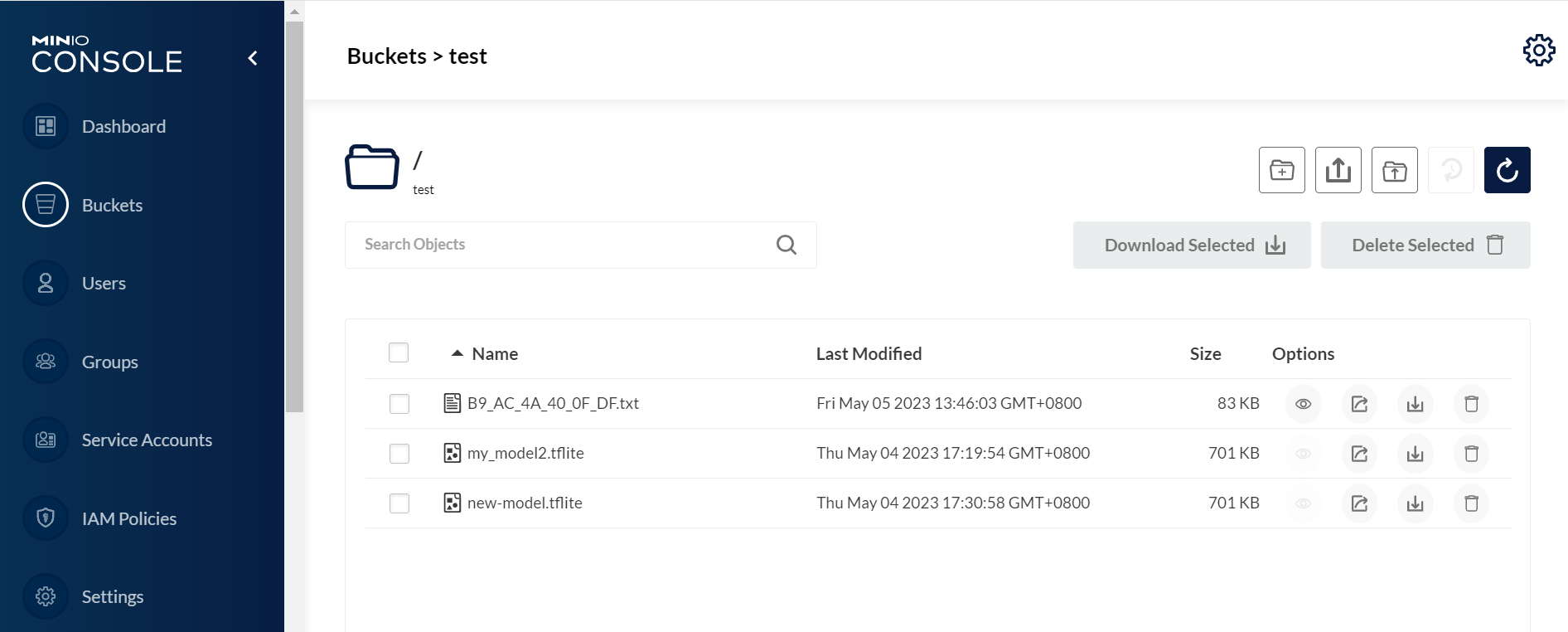}
    \caption{Minio Server}
    \label{fig:minio}
\end{figure}

The system we developed generates a large amount of unstructured data during use, such as log files of wearer walking, backup algorithm models, and so on. For the user's experience, we adopted the method of incremental training, for which we built a MinIO server in the figure \ref{fig:minio}.

MinIO is an object storage service based on Apache License v2.0 open source protocol. Compatible with the Amazon S3 cloud storage service interface, it is ideal for storing large volumes of unstructured data, such as images, videos, log files, backup data, and container/virtual machine images, and an object file can be any size, from a few kb to a maximum of 5T. MinIO is a very lightweight service that can be easily combined with other applications, such as NodeJS, Redis, or MySQL.

We used docker tools to help build minio server. After getting the minio image from the official website, we activated it according to the specified instructions, then set the user account information, and created the model folder and data folder. During the subsequent operation, the dynamic data collected by the device will be uploaded here. At the same time, with the subsequent updates of our team, there will be more detection models to provide for everyone to use.

\section{Models}
\label{sec:model}

\subsection{Fall Data acquizition}

\begin{figure}[htbp]
\centering
\subfigure[]{\includegraphics[width=0.4\linewidth]{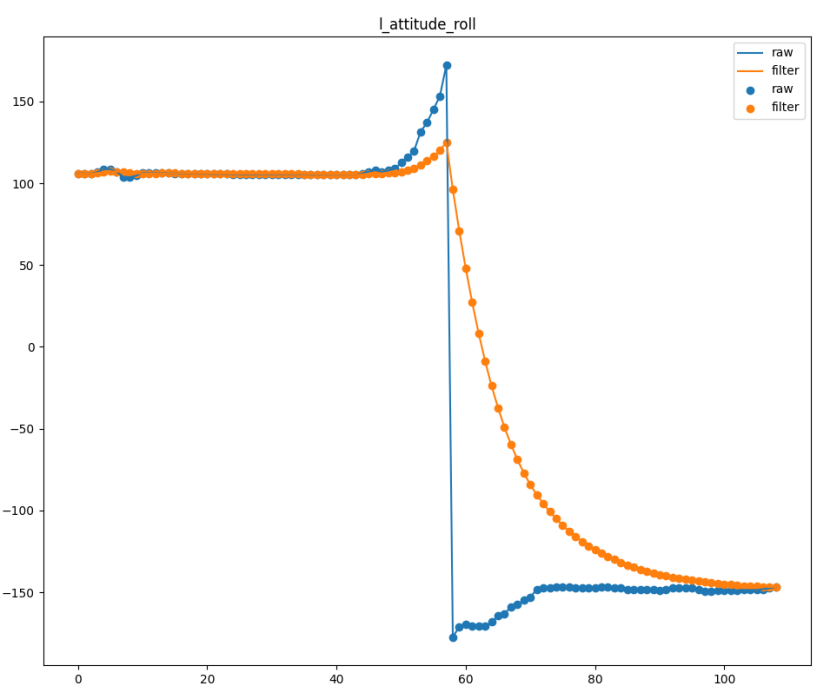}}
\hspace{0cm}
\subfigure[]{\includegraphics[width=0.4\linewidth]{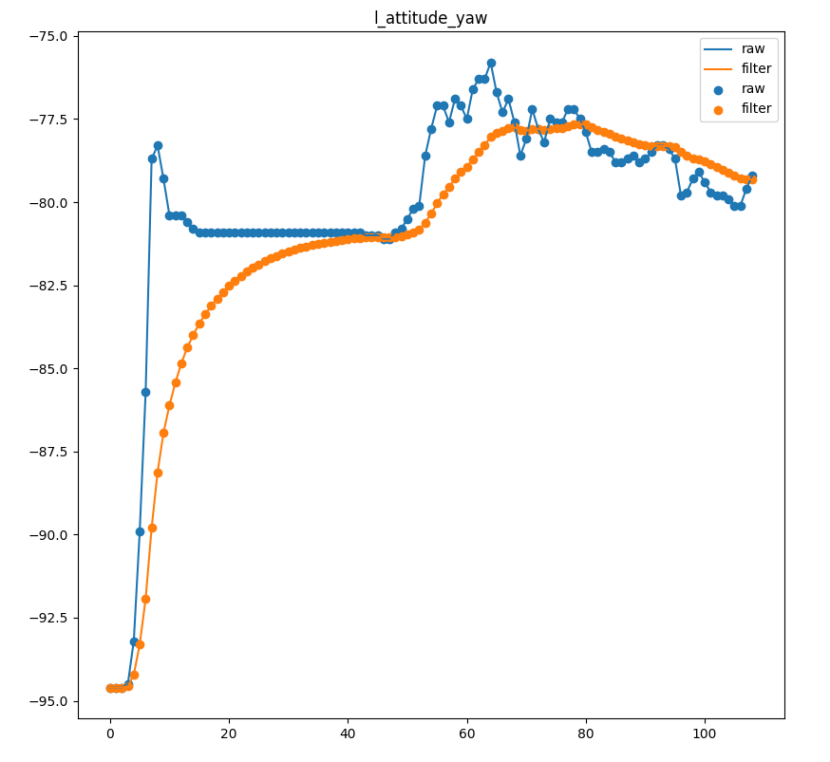}}
\vfill
\subfigure[]{\includegraphics[width=0.4\linewidth]{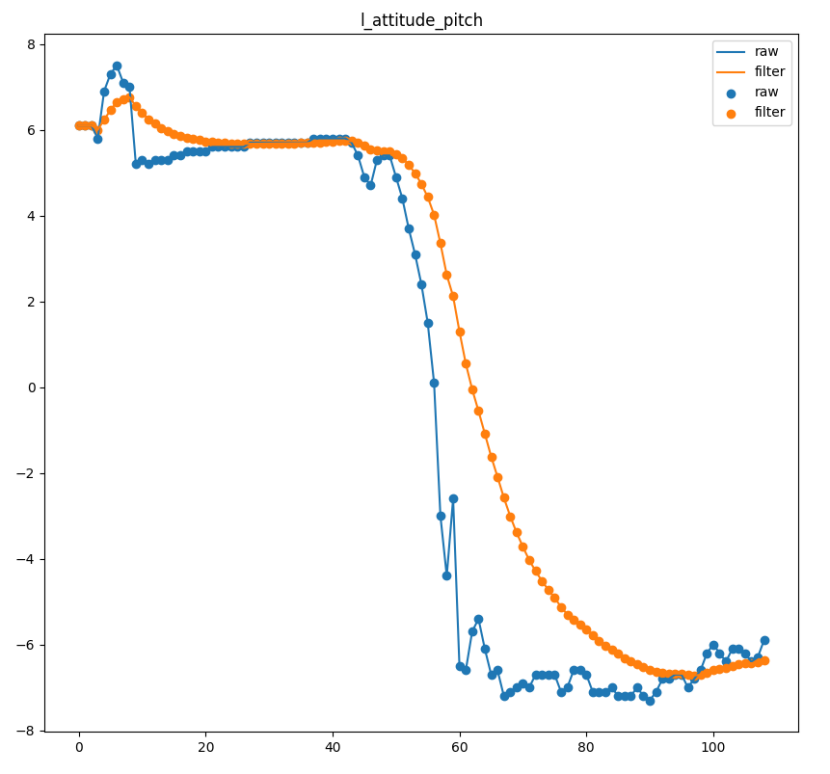}}
\hspace{0cm}
\subfigure[]{\includegraphics[width=0.4\linewidth]{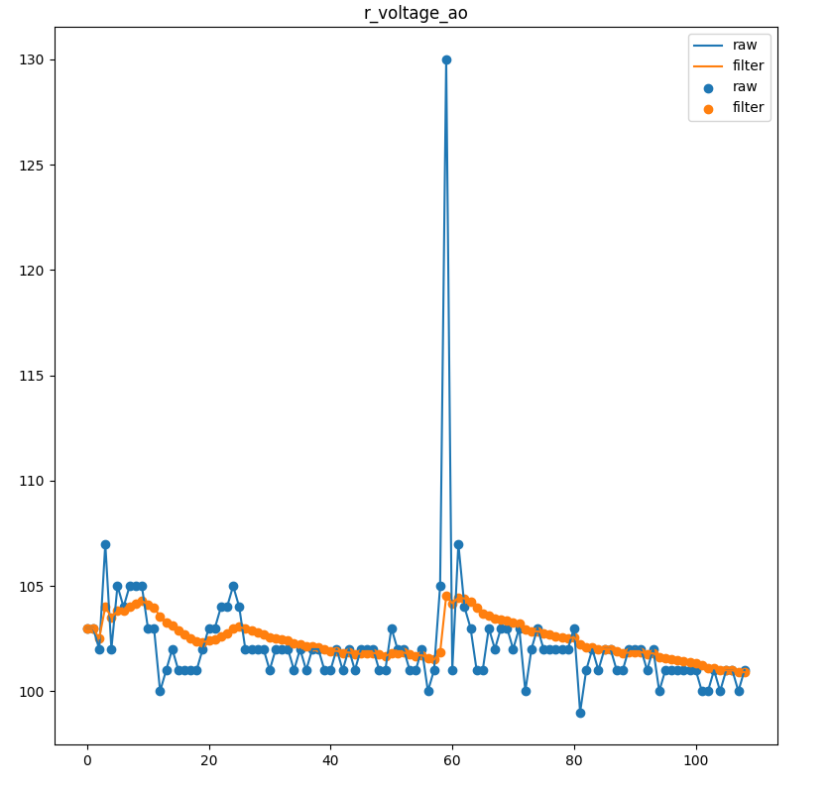}}
\caption{Plot using the Kalman filter (blue line) versus not using the Kalman filter (red line).(a)Comparison chart of before and after processing of roll angles.(b)Pitch angle before and after processing comparison chart.(c)Yaw before and after processing comparison chart.(d)Comparison chart of left foot pressure before and after processing}
\label{fig:kalman}
\end{figure}

In view of the characteristics of time series data, we uses Kalman filter for data denoising, sets the first estimate as the current value, sets the parameters A=1, W(k)=0 (Because the measurement of the parameter error of the equipment is more complicated, so we ignore this part, and this prediction is more estimated according to the trend of the overall data) and H=1.  
\begin{equation}
    X(k|k-1)=AX(k-1|k-1)+BU(k)+W(k)
\end{equation}
Figure \ref{fig:kalman} is the three-dimensional azimuth and pressure of the left foot four dimensions of the comparison chart. From (a) can be seen that the abrupt change of the azimuth angle at group 57 has been denoising into a buffer value decline; in (b), the value decline process of group 50 to group 80 is affected by noise ups and downs, and the numerical change after processing is more smooth in line with the laws of physics; (c) and (d) are the same. Kalman filter can smooth the fluctuation of abnormal data well and make the data distributed according to the trend of the whole data.
Overall, we can find that the denoising effect of the Kalman filter is very good.

Since the training data samples are relatively small, in order to improve the recognition accuracy of the training model, we can use random transformation methods to enhance the data, such as dithering, flipping, zooming in or out, bending, arranging, sliding windows, etc. These methods are the most direct ways to enhance them. In addition, neural network-based models can be used to acquire time series from feature distributions to generate new time series data. Commonly used neural networks include LSTMs and time CNNs, which can map input sequences directly to output sequences to generate samples that can be fake and authentic.

Falling or walking normally is a physical law, so the dataset has to follow that law as well. Therefore, processing methods such as dithering and flipping are not suitable, which will destroy the integrity and regularity of the data sequence.

\begin{figure}[htbp]
    \centering
    \includegraphics[scale=0.7]{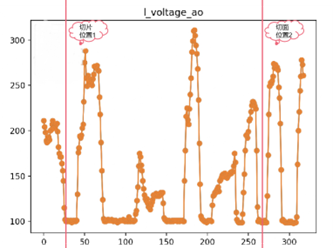}
    \caption{Schematic diagram of pre-slice data and slice locations
}
    \label{fig:stren1}
\end{figure}

\begin{figure}[htbp]
    \centering
    \includegraphics[scale=0.7]{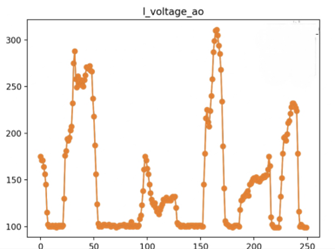}
    \caption{Schematic diagram of the data after slicing
}
    \label{fig:stren2}
\end{figure}

Therefore, slicing is mainly adopted here, which means a long data series of normal walking is divided into several segments (I did not segment the fall data here because the fall data itself is relatively short and the fluctuations of the fall data will be concentrated. Segmentation will affect the classification and judgment).
Figure \ref{fig:stren1} is the data sequence diagram on the pressure dimension of the left foot, you can see that it shows a regular undulating state. Set the first slice breakpoint in group 25 and set the second slice breakpoint in group 275. And finally we intercept a new data segment with a length of 250 as shown in figure \ref{fig:stren2}. After performing this operation on all normal walking data, the original dataset was expanded by nearly half, achieving data enhancement.

In this paper, we proposed the model named FallSeqTCNs for Fall Detection.
 
\subsection{FallSeqTCN}

\begin{figure}[htb]
    \centering
    \includegraphics[width=0.5\linewidth]{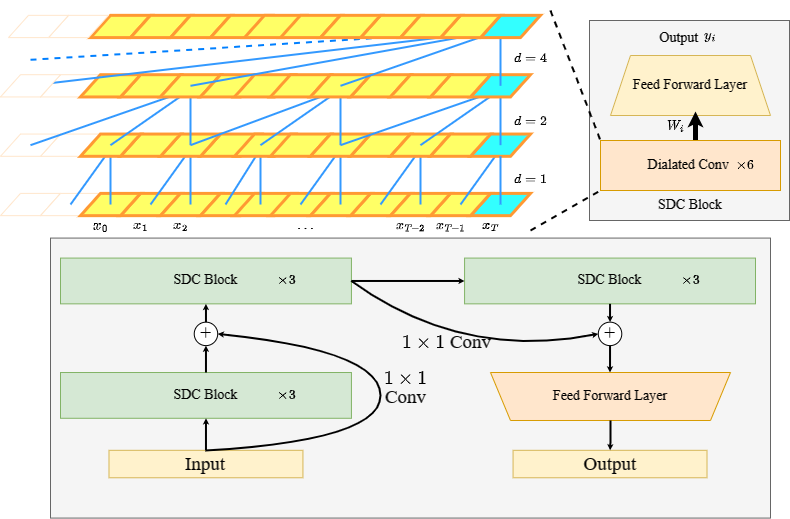}
    \caption{FallSeqTCN}
    \label{fig:FallSeqTCN}
\end{figure}

In figure \ref{fig:FallSeqTCN}, FallSeqTCN is a binary fall detection model based on Temporal Convolutional Network (TCN) \cite{2016WaveNet}. TCN is a time-series prediction network with dilated convolution, inspired by WaveNet and TCN, and it can process multiple temporal sequences simultaneously and require less time at the training time. Like TCN, residual-connected structure is used in FallSeqTCN, and dilated convolution and causal convolution are introduced in each block. 

The SDC Block is combined by several dilated 1-dimensional same-length zero-padding convolution networks and a linear feed-forward network. The residual connection is set between every 3 sequential SDC Blocks. We don't use the normalization skill because normalization may reduce the physical representational capacity of the original data. But we're still exploring the way that can further exert the physical meaning of the data and the most potential of a more general model to all sensor data.

For the FallSeqTCN model, each input sequence consists of {20} time-domain acceleration and {2} plantar pressure signals sampled at a rate of 18 Hz. Like windowing filtering, we transfer the sequence data to window batches with step 1, length 64, then fit batches into the model for training and test. The ratio of train and test is 7:3. The prediction probability is obtained from softmax layer and finally a binary fall detection result is obtained. To address overfitting issues during training, we also added optional Dropout.

\section{Experiments}

In this section, we present details of our experiment settings and the corresponding results. We conduct comprehensive analyses and investigations to illustrate the effectiveness of our FallSeqTCN model. We have provided the data and code of FallSeqTCN along with this submission.

\subsection{Datasets}
We use two datasets to evaluate the performance of FallSeqTCN:
\begin{itemize}
    \item [1)]
    \textbf{UMAFall:} is collected through the systematic emulation of a set of predefined ADLs (Activities of Daily Life) and falls in 2016. 
    \item [2)]
    \textbf{Our data:} is collected by the left and right foot data awareness device that the subjects performed 11 normal walking, 10 forward falls and 5 left side falls in the same experimental environment.
\end{itemize}

\subsection{Baselines}
We compare our FallSeqTCN with the following  models to evaluate the effectiveness of our approach:
\begin{itemize}
    \item [1)]
    \textbf{SVM:} uses SVM algorithm to get the combined acceleration, acceleration and attitude Angle thresholds of classified falls and daily behaviors, and finally reconstructs the prediction algorithm on the single chip computer to realize real-time prediction of fall behaviors. 
    \item [2)]
    \textbf{Decision Tree:} uses decision tree to build the mapping relationship between object attributes and object values, and then makes fall prediction..
    \item [3)]
    \textbf{LSTM:} is based on a single-layer long short-term memory network to make predictions, using large memory to store partial outputs of their multiple cell gates.
\end{itemize}

\subsection{Experimental Setup}

\begin{CJK}{UTF8}{gbsn}

As a binary classification problem model for normal and fall, this model evaluates the classification results by constructing a confusion matrix of the classification results of the test set. The four values in the confusion matrix are defined as T (True) for correct, F (False) for error, P (Positive) for 1, and N (Negative) for 0):
TP: The predicted value is 1, the actual value is 1, and the prediction is correct. FP: The prediction is 1, but the actual value is 0, and the prediction is wrong. FN: The predicted value is 0, but the actual value is 1, and the prediction is wrong. TN: The prediction is 0, the actual is 0, and the prediction is correct.

According to these four indicators, we can use the formula to calculate three performance indicators: Accuracy, which is the percentage of the predicted correct results in the total sample; Precision, which is the probability that all predicted positive samples will actually be positive; Recall, which is the probability that a sample is predicted to be positive in a sample that is actually positive.

\begin{equation}
    Accuracy=\frac{TP+TN}{TP+TN+FP+FN}\times100\%\label{1}
\end{equation}
\begin{equation}
    Precision=\frac{TP}{TP+FP}\times100\%\label{2}
\end{equation}
\begin{equation}
    Recall=\frac{TP}{TP+FN}\times100\%\label{3}
\end{equation}

\subsection{Results}

Through comparative training, it can be seen that the time convolutional network have a stronger ability to capture the characteristics of effective long-term series.
The Recall of SeqTCN in self-test data set and UMAFall data set reached 83\% and 85\% respectively, and its F1 scores reached 0.90 and 0.85 respectively.
\begin{itemize}
    \item [1)]
    \textbf{Overall Performance:} Through comparative training, it can be seen that the time convolutional network have a stronger ability to capture the characteristics of effective long-term series.The Recall of SeqTCN in self-test data set and UMAFall data set reached 83\% and 85\% respectively, and its F1 scores reached 0.90 and 0.85 respectively. 
    \item [2)]
    \textbf{Low Risk:} Since this is a fall detection model, recall should be increased as much as possible while the precision is reasonable. After comparison, F1 scores of TCN are higher, and the recall of both of them reaches more than 80\%, which is the reason why the TCN model is finally selected.
\end{itemize}

\end{CJK}

\begin{table}[htbp]
\caption{Model Comparison}
\begin{tabular}{l|cc|cc|cc|cc}
\toprule
\hline
\multicolumn{1}{c|}{dataset}    & UMAFall       & Our data      & UMAFall        & Our data      & UMAFall       & Our data      & UMAFall       & Our data      \\
\multicolumn{1}{c|}{model name} & \multicolumn{2}{c|}{Accuracy} & \multicolumn{2}{c|}{Precision} & \multicolumn{2}{c|}{Recall}   & \multicolumn{2}{c}{F1 Score}  \\ \hline
SVM                             & 0.76          & 0.83          & \textbf{1}     & 0.31          & 0.01          & 0.83          & 0.02          & 0.45          \\
Decision Tree                   & 0.91          & 0.91          & 0.84           & 0.1           & 0.77          & 0.50          & 0.80          & 0.16          \\
LSTM                            & 0.76          & 0.77          & \textbf{1}     & 0.25          & 0.01          & 0.83          & 0.02          & 0.38          \\
SeqTCN (Ours)                   & \textbf{0.92} & 0.98          & 0.84           & \textbf{1}    & \textbf{0.85} & \textbf{0.83} & \textbf{0.85} & {\ul 0.90}    \\ \hline
\bottomrule
\end{tabular}
\end{table}

\section{Conclusion}

This paper has presented an extensive study on motion capture based on embedded sensors, including gyroscopes, accelerometers and pressure sensors. We built a complete fall detection system named TSFallDetect. We have conducted empirical studies on existing data sets and systematically collected data sets respectively, and the results show that the model has advantages over traditional methods, which confirms the feasibility and effectiveness of our system. The time convolutional network has a strong ability to capture effective long time series features, which confirms the potential of the network for embedded sensor-based motion capture. The code is available at \url{https://github.com/WuShaoa/SensorDataClassification-TCN/tree/main/SeqClassifyCNN}.



\begin{thebibliography}{10}
\providecommand{\url}[1]{#1}
\csname url@samestyle\endcsname
\providecommand{\newblock}{\relax}
\providecommand{\bibinfo}[2]{#2}
\providecommand{\BIBentrySTDinterwordspacing}{\spaceskip=0pt\relax}
\providecommand{\BIBentryALTinterwordstretchfactor}{4}
\providecommand{\BIBentryALTinterwordspacing}{\spaceskip=\fontdimen2\font plus
\BIBentryALTinterwordstretchfactor\fontdimen3\font minus
  \fontdimen4\font\relax}
\providecommand{\BIBforeignlanguage}[2]{{%
\expandafter\ifx\csname l@#1\endcsname\relax
\typeout{** WARNING: IEEEtran.bst: No hyphenation pattern has been}%
\typeout{** loaded for the language `#1'. Using the pattern for}%
\typeout{** the default language instead.}%
\else
\language=\csname l@#1\endcsname
\fi
#2}}
\providecommand{\BIBdecl}{\relax}
\BIBdecl
\bibitem{kour2014real}
G.~Kour and R.~Saabne, ``Real-time segmentation of on-line handwritten arabic
  script,'' in \emph{Frontiers in Handwriting Recognition (ICFHR), 2014 14th
  International Conference on}.\hskip 1em plus 0.5em minus 0.4em\relax IEEE,
  2014, pp. 417--422.

\bibitem{kour2014fast}
G.~Kour and R.~Saabne, ``Fast classification of handwritten on-line arabic characters,'' in
  \emph{Soft Computing and Pattern Recognition (SoCPaR), 2014 6th International
  Conference of}.\hskip 1em plus 0.5em minus 0.4em\relax IEEE, 2014, pp.
  312--318.

\bibitem{hadash2018estimate}
G.~Hadash, E.~Kermany, B.~Carmeli, O.~Lavi, G.~Kour, and A.~Jacovi, ``Estimate
  and replace: A novel approach to integrating deep neural networks with
  existing applications,'' \emph{arXiv preprint arXiv:1804.09028}, 2018.

\bibitem{2017UMAFall}
E.~Casilari, J.~A. Santoyo-Ramón, and J.~M. Cano-García, ``Umafall: A
  multisensor dataset for the research on automatic fall detection,''
  \emph{Procedia Computer Science}, vol. 110, pp. 32--39, 2017.

\bibitem{h2w01s-17}
\BIBentryALTinterwordspacing
S.~Spinsante, E.~Gambi, L.~Montanini, D.~Perla, and A.~Del~Campo, ``Tst
  footwear-based dataset for fall detection (tst fb4fd),'' 2017. [Online].
  Available: \url{https://dx.doi.org/10.21227/H2W01S}
\BIBentrySTDinterwordspacing

\bibitem{2014A}
S.~Gasparrini, E.~Cippitelli, S.~Spinsante, and E.~Gambi, ``A depth-based fall
  detection system using a kinect® sensor,'' \emph{Sensors}, vol.~14, no.~2,
  pp. 2756--2775, 2014.

\bibitem{2017Fall}
E.~E. Stone and M.~Skubic, ``Fall detection in homes of older adults using the
  microsoft kinect,'' \emph{IEEE Journal of Biomedical \& Health Informatics},
  vol.~19, no.~1, pp. 290--301, 2017.

\bibitem{0A}
N.~Fletcher-Lloyd, A.~I. Serban, M.~Kolanko, D.~Wingfield, D.~Wilson,
  R.~Nilforooshan, P.~Barnaghi, and E.~Soreq, ``A markov chain model for
  identifying changes in daily activity patterns of people living with
  dementia,'' \emph{IEEE Internet of Things Journal}, vol.~PP.

\bibitem{2020Fall}
S.~T. Hsieh and C.~L. Lin, ``Fall detection algorithm based on mpu6050 and
  long-term short-term memory network,'' in \emph{2020 International Automatic
  Control Conference (CACS)}, 2020.

\bibitem{2016WaveNet}
A.~V.~D. Oord, S.~Dieleman, H.~Zen, K.~Simonyan, and K.~Kavukcuoglu, ``Wavenet:
  A generative model for raw audio,'' 2016.

\bibitem{9471869}
T.~Vaiyapuri, E.~L. Lydia, M.~Y. Sikkandar, V.~G. Díaz, I.~V. Pustokhina, and
  D.~A. Pustokhin, ``Internet of things and deep learning enabled elderly fall
  detection model for smart homecare,'' \emph{IEEE Access}, vol.~9, pp.
  113\,879--113\,888, 2021.

\bibitem{2020A}
E.~Casilari, R.~Lora-Rivera, and F.~García-Lagos, ``A study on the application
  of convolutional neural networks to fall detection evaluated with multiple
  public datasets,'' \emph{Sensors}, vol.~20, no.~5, p. 1466, 2020.

\bibitem{vicnn63}
K.~Adhikari, H.~Bouchachia, and H.~Nait-Charif, ``Activity recognition for
  indoor fall detection using convolutional neural network,'' in \emph{2017
  Fifteenth IAPR International Conference on Machine Vision Applications
  (MVA)}, 2017, pp. 81--84.

\bibitem{vicnn64}
X.~Li, T.~Pang, W.~Liu, and T.~Wang, ``Fall detection for elderly person care
  using convolutional neural networks,'' in \emph{2017 10th International
  Congress on Image and Signal Processing, BioMedical Engineering and
  Informatics (CISP-BMEI)}, 2017, pp. 1--6.

\bibitem{vicnn73}
N.~Lu, Y.~Wu, L.~Feng, and J.~Song, ``Deep learning for fall detection: 3d-cnn
  combined with lstm on video kinematic data,'' \emph{IEEE Journal of
  Biomedical and Health Informatics}, vol.~PP, pp. 1--1, 02 2018.

\bibitem{3DSkeleton80}
\BIBentryALTinterwordspacing
T.-H. Tsai and C.-W. Hsu, ``Implementation of fall detection system based on 3d
  skeleton for deep learning technique,'' \emph{2019 IEEE 8th Global Conference
  on Consumer Electronics (GCCE)}, pp. 389--390, 2019. [Online]. Available:
  \url{https://api.semanticscholar.org/CorpusID:211686877}
\BIBentrySTDinterwordspacing

\bibitem{CasilariPrez2020ASO77}
\BIBentryALTinterwordspacing
E.~Casilari-P{\'e}rez, R.~Lora-Rivera, and F.~Garc{\'i}a-Lagos, ``A study on
  the application of convolutional neural networks to fall detection evaluated
  with multiple public datasets,'' \emph{Sensors (Basel, Switzerland)},
  vol.~20, 2020. [Online]. Available:
  \url{https://api.semanticscholar.org/CorpusID:212666988}
\BIBentrySTDinterwordspacing

\bibitem{cnnlstmIoTEnabled99}
\BIBentryALTinterwordspacing
J.~Xu, Z.~He, and Y.~Zhang, ``Cnn-lstm combined network for iot enabled fall
  detection applications,'' \emph{Journal of Physics: Conference Series}, vol.
  1267, no.~1, p. 012044, jul 2019. [Online]. Available:
  \url{https://dx.doi.org/10.1088/1742-6596/1267/1/012044}
\BIBentrySTDinterwordspacing

\bibitem{Torti2018EmbeddedRF105}
\BIBentryALTinterwordspacing
E.~Torti, A.~Fontanella, M.~Musci, N.~Blago, D.~P. Pau, F.~Leporati, and
  M.~Piastra, ``Embedded real-time fall detection with deep learning on
  wearable devices,'' \emph{2018 21st Euromicro Conference on Digital System
  Design (DSD)}, pp. 405--412, 2018. [Online]. Available:
  \url{https://api.semanticscholar.org/CorpusID:52985178}
\BIBentrySTDinterwordspacing

\bibitem{Theodoridis2018HumanFD106}
\BIBentryALTinterwordspacing
T.~Theodoridis, V.~Solachidis, N.~Vretos, and P.~Daras, ``Human fall detection
  from acceleration measurements using a recurrent neural network,'' 2018.
  [Online]. Available: \url{https://api.semanticscholar.org/CorpusID:196017607}
\BIBentrySTDinterwordspacing

\bibitem{2014Survey}
Y.~Delahoz and M.~Labrador, ``Survey on fall detection and fall prevention
  using wearable and external sensors,'' \emph{Sensors}, vol.~14, no.~10, p.
  19806, 2014.

\bibitem{Kabalan2017From}
Kabalan, Chaccour, Rony, Darazi, Amir, Hajjam, El, Hassani, Emmanuel, and
  Andrès, ``From fall detection to fall prevention: A generic classification
  of fall-related systems,'' \emph{IEEE Sensors Journal}, 2017.

\bibitem{2017Radar}
E.~Cippitelli, F.~Fioranelli, E.~Gambi, and S.~Spinsante, ``Radar and rgb-depth
  sensors for fall detection: A review,'' \emph{IEEE Sensors Journal}, pp.
  3585--3604, 2017.

\bibitem{2015An}
M.~S. Khan, M.~Yu, P.~Feng, L.~Wang, and J.~Chambers, ``An unsupervised
  acoustic fall detection system using source separation for sound interference
  suppression,'' \emph{Signal Processing}, vol. 110, no.~C, pp. 199--210, 2015.

\bibitem{2016Floor}
G.~Feng, J.~Mai, Z.~Ban, X.~Guo, and G.~Wang, ``Floor pressure imaging for fall
  detection with fiber-optic sensors,'' \emph{IEEE Pervasive Computing},
  vol.~15, no.~2, pp. 40--47, 2016.

\bibitem{2017Wearables}
A.~G.~A. B, ``Wearables for independent living in older adults: Gait and
  falls,'' \emph{Maturitas}, vol. 100, pp. 16--26, 2017.

\bibitem{Mukhopadhyay2014Wearable}
Mukhopadhyay and S.~Chandra, ``Wearable sensors for human activity monitoring:
  A review,'' \emph{IEEE Sensors Journal}, vol.~15, no.~3, pp. 1321--1330,
  2014.

\bibitem{2014Detecting}
A.~\"Ozdemir and B.~Barshan, ``Detecting falls with wearable sensors using
  machine learning techniques.'' \emph{Sensors}, vol.~14, no.~6, p. 10691,
  2014.

\bibitem{2016Investigation}
P.~Ntanasis, E.~Pippa, A.~T. Zdemir, B.~Barshan, and V.~Megalooikonomou,
  ``Investigation of sensor placement for accurate fall detection,'' in
  \emph{International Conference on Wireless Mobile Communication and
  Healthcare}, 2016.

\bibitem{2016An}
O.~Ahmet, ``An analysis on sensor locations of the human body for wearable fall
  detection devices: Principles and practice,'' \emph{Sensors (Basel,
  Switzerland)}, vol.~16, no.~8, 2016.

\bibitem{2013Fall}
N.~El-Bendary, Q.~Tan, F.~C. Pivot, and A.~Lam, ``Fall detection and prevention
  for the elderly: A review of trends and challenges,'' \emph{International
  Journal on Smart Sensing \& Intelligent Systems}, vol.~6, no.~3, pp.
  1230--1266, 2013.

\bibitem{2007Falls}
L.~Day, ``Falls in older people: Risk factors and strategies for prevention.''
  \emph{age \& ageing}, 2007.

\bibitem{who2023}
W.~H. Organization, 2023, \url{https://www.who.int/health-topics/ageing/}, Last
  accessed on 2023-10-25.

\end{thebibliography}

\end{document}